\documentclass[conference]{IEEEtran}
\IEEEoverridecommandlockouts

\usepackage{cite}
\usepackage{amsmath,amssymb,amsfonts}
\usepackage{algpseudocode}
\usepackage{graphicx}
\usepackage{textcomp}
\usepackage{xcolor}
\usepackage{bm}
\usepackage{psfrag}
\usepackage[utf8]{inputenc} 
\usepackage[T1]{fontenc}    
\usepackage{hyperref}       
\usepackage{url}            
\usepackage{booktabs}       
\usepackage{amsfonts}       
\usepackage{nicefrac}       
\usepackage{microtype}      
\usepackage{amsmath}
\usepackage[inline]{enumitem}
\DeclareGraphicsExtensions{.pdf,.jpeg,.png,  .eps}
\usepackage{psfrag}
\usepackage{epstopdf}
\usepackage{amsfonts}
\usepackage{mathrsfs}
\usepackage{pifont}
\usepackage{verbatim}
\usepackage{upgreek}
\usepackage{soul,color}
\usepackage{caption}

\newcommand{\Diag}{\mbox{\boldmath\bf Diag}\, }

\usepackage{algorithm}
\usepackage{mdwmath}
\usepackage{mdwtab}
\usepackage{amssymb,amsmath}
\usepackage{amsmath}
\usepackage{amsthm}
\usepackage{epstopdf}
\usepackage{tikz}
\usepackage{scalefnt}
\usepackage{subfigure}
\usepackage{MnSymbol}
\usepackage{setspace}
\usepackage{mathtools}
\usepackage{psfrag}
\usepackage{cite}
\usepackage{romannum}
\def\BibTeX{{\rm B\kern-.05em{\sc i\kern-.025em b}\kern-.08em
    T\kern-.1667em\lower.7ex\hbox{E}\kern-.125emX}}
\begin{document}

\title{{Machine Learning-based Methods for Reconfigurable Antenna Mode Selection in MIMO Systems}}

\author{\IEEEauthorblockN{Yasaman Abdollahian\IEEEauthorrefmark{1},
Ehsan Tohidi\IEEEauthorrefmark{1},  Martin~Kasparick\IEEEauthorrefmark{1},\\ Li Wang\IEEEauthorrefmark{2}, Ahmet Hasim Gokceoglu\IEEEauthorrefmark{2},  and
Slawomir Stanczak\IEEEauthorrefmark{1}} 
\IEEEauthorblockA{\IEEEauthorrefmark{1}Fraunhofer Heinrich Hertz Institute,
Berlin, Germany,\\ \IEEEauthorrefmark{2}Huawei Technologies,
Stockholm, Sweden\\
Email: yasaman.abdollahian@hhi.fraunhofer.de, ehsan.tohidi@hhi.fraunhofer.de, martin.kasparick@hhi.fraunhofer.de}}

\maketitle

\begin{abstract}
MIMO technology has enabled spatial multiple access and has provided a higher system spectral efficiency (SE). However, this technology has some drawbacks, such as the high number of RF chains that increases complexity in the system. One of the solutions to this problem can be to employ reconfigurable antennas (RAs) that can support different radiation patterns during transmission to provide similar performance with fewer RF chains. In this regard, the system aims to maximize the SE with respect to optimum beamforming design and RA mode selection. Due to the non-convexity of this problem, we propose machine learning-based methods for RA antenna mode selection in both dynamic and static scenarios. In the static scenario, we present how to solve the RA mode selection problem, an integer optimization problem in nature, via deep convolutional neural networks (DCNN). A Multi-Armed-bandit (MAB) consisting of offline and online training is employed for the dynamic RA state selection. For the proposed MAB, the computational complexity of the optimization problem is reduced. Finally, the proposed methods in both dynamic and static scenarios are compared with exhaustive search and random selection methods.
\end{abstract}


\section{Introduction}
In wireless communication, one way to increase the SE is by employing MIMO technology. Needless to say, increasing the number of antennas in MIMO raises new challenges, such as the higher complexity of hardware, e.g., a higher number of RF chains, and software \cite{marzetta2016fundamentals, 9095425, 8537943}. RA has been known as a feasible solution to reduce the number of RF chains. RA can be implemented with different procedures, such as modifications in radiation patterns, antenna polarizations, frequency, or combinations of them \cite{costantine2015reconfigurable}, \cite{bahceci2016efficient}. This reconfigurability can be done with the asset of RF switches, which leads to changes in impedance and antenna radiation properties \cite{haupt2013reconfigurable}. Here, we concentrate on RAs that are equipped with different radiation patterns. When employing RA in wireless systems, algorithms for RA state selection that maximize the channel capacity are required. However, finding an optimal RA state is an NP-hard problem. Therefore, several studies concentrate on the RA state selection problem \cite{li2018joint, zhai2017joint}.

\subsection{Related Works}
Recently, machine learning-based methods in wireless communications are accumulating more interests in different areas, such as MIMO channel denoising, beam selection, cognitive radio, etc. Moreover, they can be employed to tackle complex and non-convex problems with lower complexity. Also, they can be implemented on graphics processing units (GPUs) to reach the solution in a more energy-efficient manner \cite{huang2019deep}.

This work \cite{vu2021machine} presents a joint antenna selection and beamforming design in which a semi-definite optimization algorithm for this non-convex problem is solved through supervised training of a neural network model. 
The authors in \cite{elbir2019joint} present a two-step algorithm for antenna selection and beamforming design that are implemented as classification and prediction problems. The antenna selection task is carried out in the first step, and the beamforming matrix is designed in the second step. The proposed ML-based method for both tasks is convolutional neural networks (CNN).
The authors in \cite{lin2019beamforming} propose a deep learning-based method for designing the analog beamforming matrix for mm-wave multiple input single output (MISO). The proposed network training is unsupervised in which, during the training, the SE of the system is calculated via an analog beamformer matrix that is the output of the network.
A comparison between conventional supervised and unsupervised deep learning methods for beamforming design is given in \cite{huang2019deep}. In the supervised case, the loss function is the mean squared error (MSE) between the predicted labels and ground truth ones. For the unsupervised case, the SE is defined as the loss function, combined with an estimated covariance matrix for each user. In \cite{lee2018deep}, deep power control for a wireless communication system is proposed based on CNNs. In this work, both supervised and unsupervised learning approaches are used to train the neural network to achieve a higher spectral or energy efficiency with a reduced training time. The work in \cite{liu2022deep} uses an unsupervised learning residual neural network (ResNet) method for jointly designing the hybrid beamforming and antenna selection in the downlink MIMO. The training of the system consists of three main parts; in the first part, by considering the fixed antenna selection matrix, the beamforming matrix is predicted. In the second part,  the extracted beamforming matrix is used to train the model for the antenna selection task, and in the last part, both systems are trained jointly. The mutual goal of the whole system is maximizing the mutual information that is calculated through network training.
 As for the dynamic RA mode selection in \cite{gulati2013learning}, a reinforcement learning (RL) based method, namely Multi-Armed-bandit (MAB), is proposed. The proposed MAB policy is upper confidence bound (UCB), in which the RA states are selected from the five possible states. As \cite{zhao2019online} shows, the UCB policy does not achieve satisfactory performance in the case of having a larger number of states in comparison to the Thompson Sampling (TS). As a result, by increasing the number of RA states, the problem becomes more complicated, and for that, in \cite{zhao2019fast}, a hierarchical TS (HTS) policy is proposed to select the best RA states with the help of clustering methods. In this model, the RA states are divided into a number of clusters, and in each cluster, the HTS is executed to extract the optimal RA states. This algorithm convergence faster in comparison to the MAB-TS and UCB with less computational complexity. The paper \cite{zhao2021online} employs MAB for RA state selection plus channel prediction for unexplored states in the case of using both policies of UCB and TS.\\

 \subsection{Contributions and Novelties}
We study the performance of a multi-user
MIMO downlink system via a joint design of RA mode selection and digital beamforming to enhance the system performance in both static and dynamic MIMO scenarios. Our contributions
are as follows: In the static scenario, $(i)$ we design a deep convolutional neural network (DCNN) with an unsupervised learning approach for RA mode selection. For this model, we propose a customized activation function for the integer optimization problem of selected RA modes, plus a customized loss function for SE calculation, and $(ii)$ we design a DCNN as the combination of transfer learning and a semi-supervised learning approach for RA mode selection. This model reduces the network training time and boosts the system's performance. In the dynamic scenario, $(iii)$ we apply MAB with different policies. For that, we include modifications to TS and UCB policies to adapt them to our problem, and $(iv)$ we apply offline training for MAB models to reduce the complexity of RA state selection. This means that both feature reduction and clustering methods are employed to find the representative RA states for the online training.


\subsection{Outline}
The rest of the paper is organized as follows. Section \Romannum{2} defines the proposed system model for the joint design of RA mode selection and beamforming matrices, followed by the problem formulation. The static and dynamic RA mode selections are described in Section \Romannum{3} and \Romannum{4}, respectively. Simulation results are presented in Section \Romannum{5}. Finally, conclusions are given in Section \Romannum{6}.

\section{System Model and Problem Formulation} \label{sys_model}
We consider the downlink of a single-cell MIMO system.
The BS is equipped with a uniform planar array antenna (UPA) with RA technology that has $N_T$ elements with $N_P$ different radiation patterns (modes). The BS is equipped with $N_{RF}$, RF chains to transmit $N_S$ data streams. The RX side is equipped with $N_R$ antennas for each user. In this paper, we consider both single (SU) and multi-user (MU) scenarios. At the BS, the transmitted symbols are processed by a digital beamformer $\boldsymbol{F}_{BB} \in \mathbb{C}^{N_{RF}\times N_S}$, and then pass through $\boldsymbol{F}_{RF} \in \mathbb{C}^{N_{T}\times N_{RF}}$. This RF precoder determines the connection of $N_T$ antennas to the $N_{RF}$ RF chains, which are considered in this paper without any phase-shifting operations. The transmit signal $\boldsymbol{x} \in \mathbb{C}^{N_{T}\times 1}$ can be defined with respect to the power-constrained beamformers as:

\begin{equation}
    \boldsymbol{x}= \boldsymbol{F}_{RF}\boldsymbol{F}_{BB}\boldsymbol{s},
\end{equation}
where $\boldsymbol{s}\in \mathbb{C}^{N_{s} \times 1}$ is the vector of transmitted symbols. The spectral efficiency of the system for each user $k= 1,2,\cdots,K$ at each OFDM subcarrier $f \in \{1, \cdots, N_f\}$, and each time step $t$ can be defined as:

\begin{align}
        R(\boldsymbol{F}_{RF},&\boldsymbol{F}_{BB},\boldsymbol{\nu}) = \sum_{k=1}^K\log_2(|\boldsymbol{I}+\frac{\rho}{N_s}\boldsymbol{C}^{-1}\boldsymbol{H}_k[f,t,\boldsymbol {\nu}]\boldsymbol{F}_{RF}[t]\notag\\ 
        & \boldsymbol{F}_{BB,k}[f,t]
        \boldsymbol{F}_{BB,k}^H[f,t]\boldsymbol{F}_{RF}^H[t]\boldsymbol{H}_k^H[f,t,\boldsymbol{\nu}]|),
        \label{eq:SE}
    \end{align}
where $\boldsymbol{C} = {\mathbf{H}}_k[f,t]{\boldsymbol{F}}_{\text{RF}}[t] \sum_{\bar{k}=1,\bar{k}\neq k}^K {\boldsymbol{F}}_{\text{BB},\bar{k}}[f,t]{\boldsymbol{F}}_{\text{BB},\bar{k}}^H[f,t]\\ {\boldsymbol{F}}_{\text{RF}}^H[t] {\mathbf{H}}_k^H[f,t] + \sigma_n^2 \boldsymbol{I}$ is the covariance matrix of both interference and noise. We denote each RA state by $\boldsymbol{\nu} = [\nu_1, \cdots, \nu_T] \in \mathcal{M} ^ {N_T \times 1}$ which contains the modes of $N_T$ antennas.  The general formulation of SE including all subcarriers and time steps is presented in \eqref{eq:SE}. However, for the sake of simplifying the illustration, we focus on one subcarrier with one and multiple time steps for static and dynamic scenarios, respectively.
The main aim of the system is to maximize the SE through an optimal design of the beamforming matrix and RA mode selection. 
 
\section{Static RA Mode Selection}
\begin{figure}
    \centering
    \includegraphics[width=0.48\textwidth]{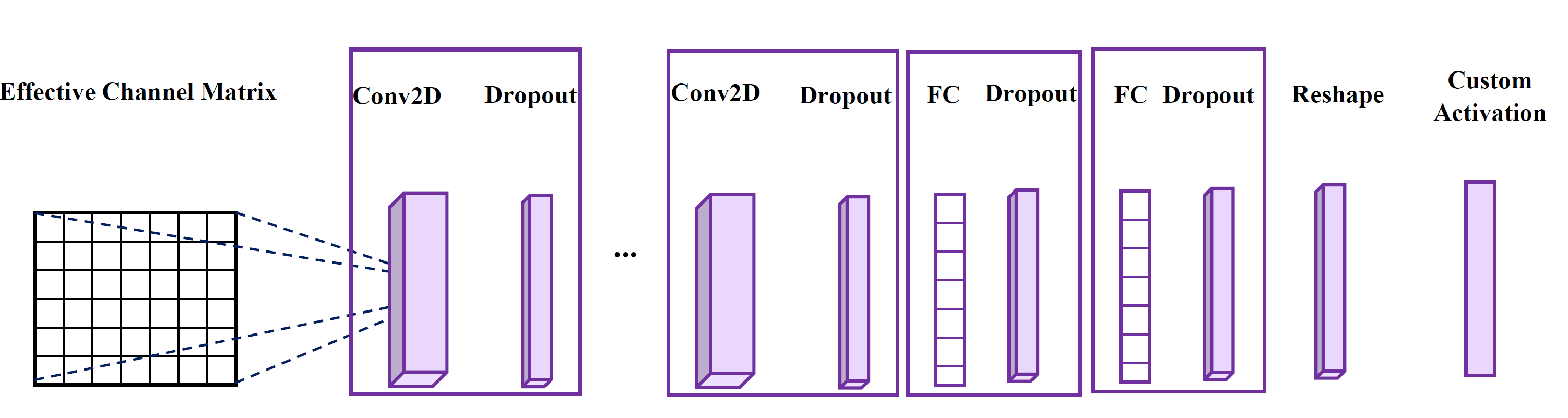}
    \caption{DCNN architecture in the proposed scheme }
    \label{fig:DCNN}
\end{figure}

The RA state selection in static scenarios can be defined via a binary decision-making matrix $\boldsymbol{W} \in \{0,1\}^{N_T \times N_P}$, meaning that if $W_{m,\nu}=1$, the $m$th antenna has selected mode $\nu$ and $0$ otherwise. Therefore, the channel matrix after RA mode selection can be defined as follows:
\begin{equation}
    \begin{aligned}
        \boldsymbol{H} [\boldsymbol{W} ] &= \sum_{\nu=1}^{N_P} \boldsymbol{H}_s[\nu] \Diag(\boldsymbol{W}_{\nu})\\
        \text{s.t.} & \sum_{\nu=1}^{N_P} W_{m,\nu} = 1, ~\forall m\\
        & \boldsymbol{W}\in\{0,1\}^{N_T\times N_P}
        \label{eq:RAMat}
    \end{aligned}
\end{equation}
where $\Diag(\cdot)$ defines the diagonal matrix operator. $\boldsymbol{W}_{\nu}$ denotes the $\nu$th column of $\boldsymbol{W}$, and $\boldsymbol{H}_s[\nu]$ defines the channel when all transmit antennas select the specific mode $\nu$.
We provide a model in which deep learning can be adopted in MIMO for accomplishing the joint problem of RA mode selection and digital beamforming design.  The proposed DCNN model is depicted in Fig.\ref{fig:DCNN}. With respect to the convolutional operators in the proposed DCNN model, the spatial features of the channel are extracted to capture the beneficial properties of digital precoding and RA mode selection matrices. These convolutional operators are mainly applicable to image-processing tasks. As a result, we transfer the channel coefficient matrix in image shape to apply convolutional layers on it. Here, the input layer contains the effective channel matrix in which its real and imaginary parts are separated and stacked as a 2D image. To extract the RA mode selection matrix in the last layer, we propose a customized \texttt{softmax} activation function in order to choose the optimum RA modes for each RF chain. In other words, the proposed customized activation function \eqref{eq:softmax} tries to find the suitable RA mode selection matrix as defined in \eqref{eq:RAMat}. This means that it shifts the output to one-hot encoder variables by increasing $\alpha$ through each epoch's training.
\begin{equation}
    f(x)= \frac{ \exp{(\alpha x)}}{\sum_{i=1}^{n} \exp{(\alpha x)}}
    \label{eq:softmax}
\end{equation}
\subsection{Learning Approach}
We consider two learning approaches for the proposed DCNN, namely unsupervised and semi-supervised learning. It should be noted that applying unsupervised learning for DCNN in multi-user scenarios requires more training time and a deeper network. 
Therefore, to reduce the training time, we attempt to apply semi-supervised learning in combination with transfer learning in the second approach.
\subsubsection{Unsupervised Learning}
Regarding this learning approach, we customize the loss function to  {solve \eqref{eq:SE}}.  For the multi-user scenario, we design the digital beamforming matrix via \cite{spencer2004zero}, which eliminates the users' interference by finding the null space of the right singular vector of the users' channels. Following the beamforming design, the SE of the system is calculated according to \eqref{eq:SE}. During each epoch's training, the DCNN attempts to minimize the negative of SE with respect to the predefined optimizer method. 
\subsubsection{Semi-Supervised Learning}
We adopt the idea of combining semi-supervised and transfer learning in \cite{lee2018deep}. We propose to use a transfer learning method so that, in the first step, we train the DCNN network in a supervised manner and then use the trained DCNN model for the unsupervised task. Thereby, less time is needed for training the main task, which is the RA mode selection with unsupervised learning. For the supervised DCNN, the RA modes are extracted via the exhaustive search method and transferred to DCNN as ground truth labels. During learning, the binary cross entropy loss \cite{ruby2020binary} between ground truth and predicted labels are minimized. After initialization of the weights, biases, and training, the network is used for unsupervised learning. We use the same architecture for semi-supervised and unsupervised DCNN, where only the learning procedure is changed.

\section{Dynamic RA Mode Selection}
In this section, we assume a dynamic scenario with a mobile UE. A moving UE results in a time-varying channel. As a result, the RA state selection is more challenging compared to the static scenario. Dynamic RA mode selection can be accounted as an online mode selection, meaning that in each time step, the RA mode has to be updated. It should be noted that the problem of RA mode selection and RA state selection are equivalent. 

We aim to develop a low-complex algorithm consisting of offline and online training. The offline training comprises the two steps of feature reduction and clustering methods. During offline training, we extract the important features of the effective channel via the PCA method \cite{bishop2006pattern}. In the next step, we apply the K-means algorithm as a clustering method \cite{bishop2006pattern} to group similar RA states. In other words, by choosing an appropriate number of clusters, the within cluster's sum of squares distances (WCSS) will be negligible. Following that, the representative clusters' points (the points that have lower WCSS) are extracted to be used in the next step, which is online training. Thus, the representative points obtained in the offline training procedure could be considered as the states of the dynamic scenario, in which we employ MAB. For MAB, similar to other RL methods, we need to assign a policy and derive the rewards and punishments of states from that. Here, we modify two 
policies of UCB \cite{auer2002finite} and TS \cite{agrawal2012analysis}. The main responsibility of MAB is to provide a balance between exploration and exploitation (EE) for arms selection. For our problem, the RA states are considered as the MAB arms, and each arm has a reward that is the SE of the system. The aim is to select the arm that leads the system to a higher reward. 
\subsection{MAB Policies}
\subsubsection{UCB}
 This policy requires the average of all instantaneous reward values up to the present time step named as $\bar{R_i}(t)$. 
 Moreover, UCB needs the number of times up to the current time step that each state has been selected $n_i(t)$. These two values are updated regarding each arm and time step as follows:
\begin{equation}
    \bar{R_i}(t) = \begin{cases} \frac{\bar{R_i}(t-1)n_i(t-1)+ R_i(t) }{n_i(t-1) +1}  & \text{if state $i$ is selected }\\
    \bar{R_i}(t-1) & \text{otherwise}
    \label{eq:r}
    \end{cases}
\end{equation}
and
\begin{equation}
    n_i(t) = \begin{cases} n_i(t-1)+1  & \text{if state $i$ is selected }\\
    n_i(t-1) & \text{otherwise}
    \label{eq:n}
    \end{cases}
\end{equation}
Here, we customize the UCB algorithm to provide a balance between EE for our problem. This balancing is defined via the $\gamma$ factor, which is an arbitrary positive real number, applied to the exploitation term of EE.
\begin{algorithm}
\caption{UCB Policy \label{alg:UCB}\cite{gulati2013learning}}
\begin{algorithmic}[1]
\State  $n_i=0$, $\bar{R_i}=0$;

\State Select each state at least once and update $n_i, \bar{R_i}$
\For{ $t=1$ to $N_{\text{{time step}}}$} 
\State Select a RA state that maximizes $\gamma\bar{R_i}+\sqrt{\frac{2\ln{(t})}{n_i}}$
\State Update $\bar{R_i}$, $n_i$ for state i, w.r.t to \eqref{eq:r}, \eqref{eq:n}
\EndFor
\end{algorithmic}
\end{algorithm}
\subsubsection{TS}
This policy assumes that the mean reward of each arm has a particular distribution; typically, the Beta distribution is considered. At each time step, an arm from the Beta distribution is selected, and if the reward regarding that arm is larger than the Bernoulli reward, the success probability of that state is updated; otherwise, the failure probability is increased.

Regarding the TS policy, we propose four terms for EE tradeoff balancing. These terms are $\lambda$, $\omega$, the scalar values from real positive numbers for Beta distribution parameters, and $\Delta S$, $\Delta F$ for success and failure updates of the Beta distribution. The $S_i$ and $F_i$ parameters represent the number of successes and failures of each state.
\begin{algorithm}
\caption{Thompson Sampling Policy \label{alg:TS}\cite{zhao2021online}}
\begin{algorithmic}[1]
\State $S_i =0$, $F_i =0$;
\For{ $t=1$ to $N_{\text{{time step}}}$} 
    \State For each arm $i=1,\ldots, N$ sample $\eta_i(t)$ from Beta$(S_{i}+\lambda, F_{i}+\omega)$ distribution.
    \State Play arm $i(t):= \arg\max \eta_i(t)$ and observe reward $\Tilde{r_t}$ 
    \State Perform Bernouli trail with probability $\Tilde{r_t} \in [0,1]$ and observe output $r_t \in \{1,0\}$
    \If {$r_t=1$} 
    \State $S_i= S_{i}+ \Delta S$
    \Else 
     \State {$F_i = F_{i}+\Delta F$}
    
    \EndIf
\EndFor
\end{algorithmic}
\end{algorithm}
\section{Simulation Results}
In this section, we evaluate the performance of the proposed methods, that is, DCNN for RA mode selection in static scenarios, compared with an exhaustive search method, and MAB with mentioned policies for dynamic scenarios.
\subsection{Static Scenario}
Regarding the DCNN implementation, we use TensorFlow Keras in Python. We consider QuaDRiGa channels  \cite{jaeckel2014quadriga} for our model.
For the static scenario, we generate $20000$ samples in which UEs are positioned uniformly distributed in $50-100m$. The SE is evaluated as  a performance metric. For the single-user case, the proposed DCNN model consists of six Conv2D layers, each three of which have a filter size of $64$ and $128$ with a kernel size of $2 \times 2$, respectively. Then, after the flattening layer, two pairs of fully connected (FC) and dropout layers with the size of $512$ and $0.1$ are used. For all layers, we use the \texttt{ReLU} activation function \cite{agarap2018deep}, and for the output layer, we apply the customized activation \eqref{eq:softmax}. The number of epochs and batch size are $200$ and $1024$, respectively. 

Regarding the DCNN architecture for the two-user scenario, we consider nine Conv2D layers, each three of which have a filter size of $32$, $64$, and $128$, respectively, with a kernel size of $2 \times 2$. We use the \texttt{tanh} as a non-linear activation function for all layers besides the output layer that is equipped with the customized function. Next, the extracted features matrix is flattened and followed by a pair of fully FC layers with $128$ units. Finally, a dropout layer with the rate of $0.1$ for decision-making and training regularization is considered. The number of epochs and batch sizes are $50$ and $64$, {respectively.}

\begin{figure}
\centering
\subfigure[]{\label{fig:a}\includegraphics[width=0.24\textwidth]{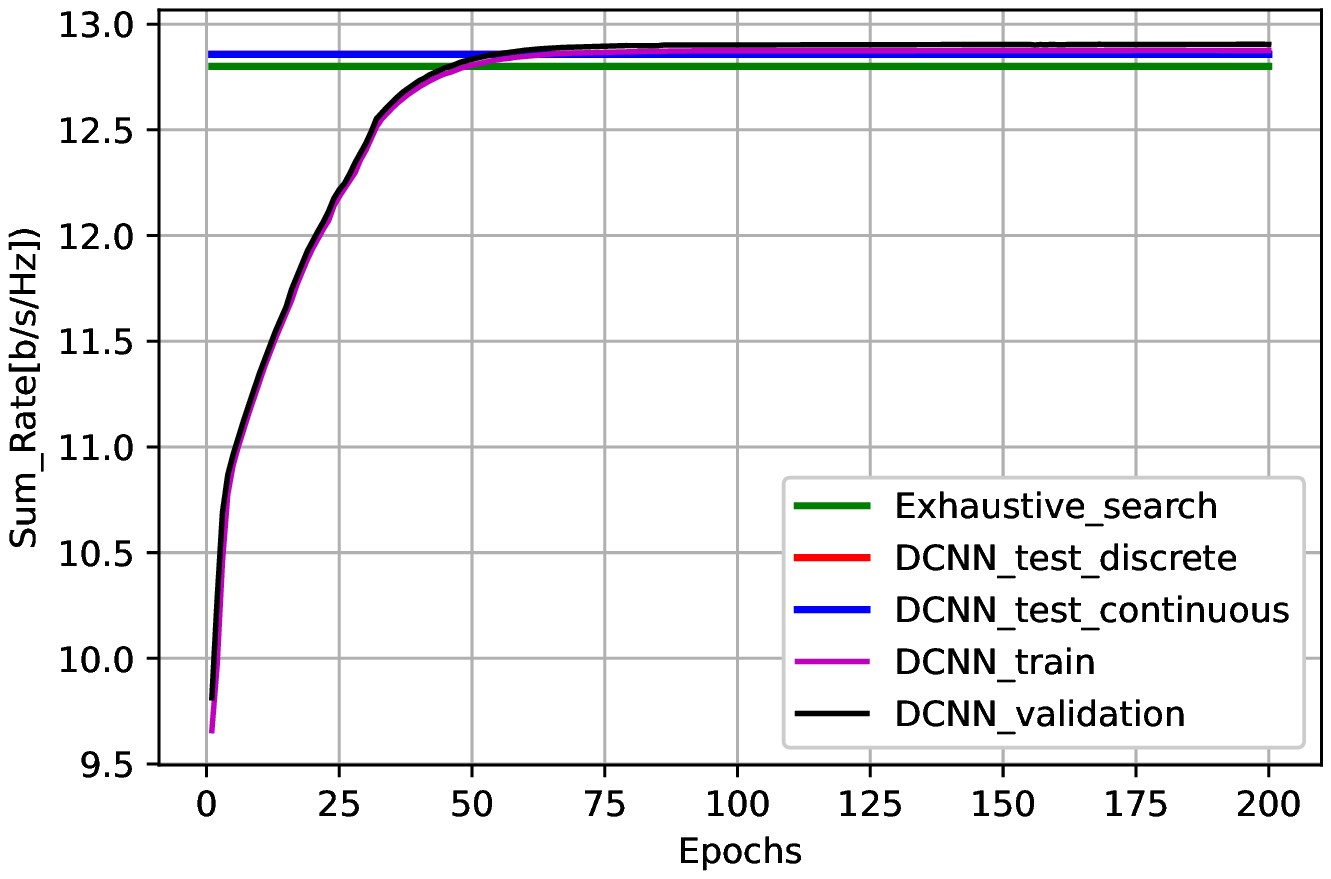}} 
\subfigure[]{\label{fig:b}\includegraphics[width=0.24\textwidth]{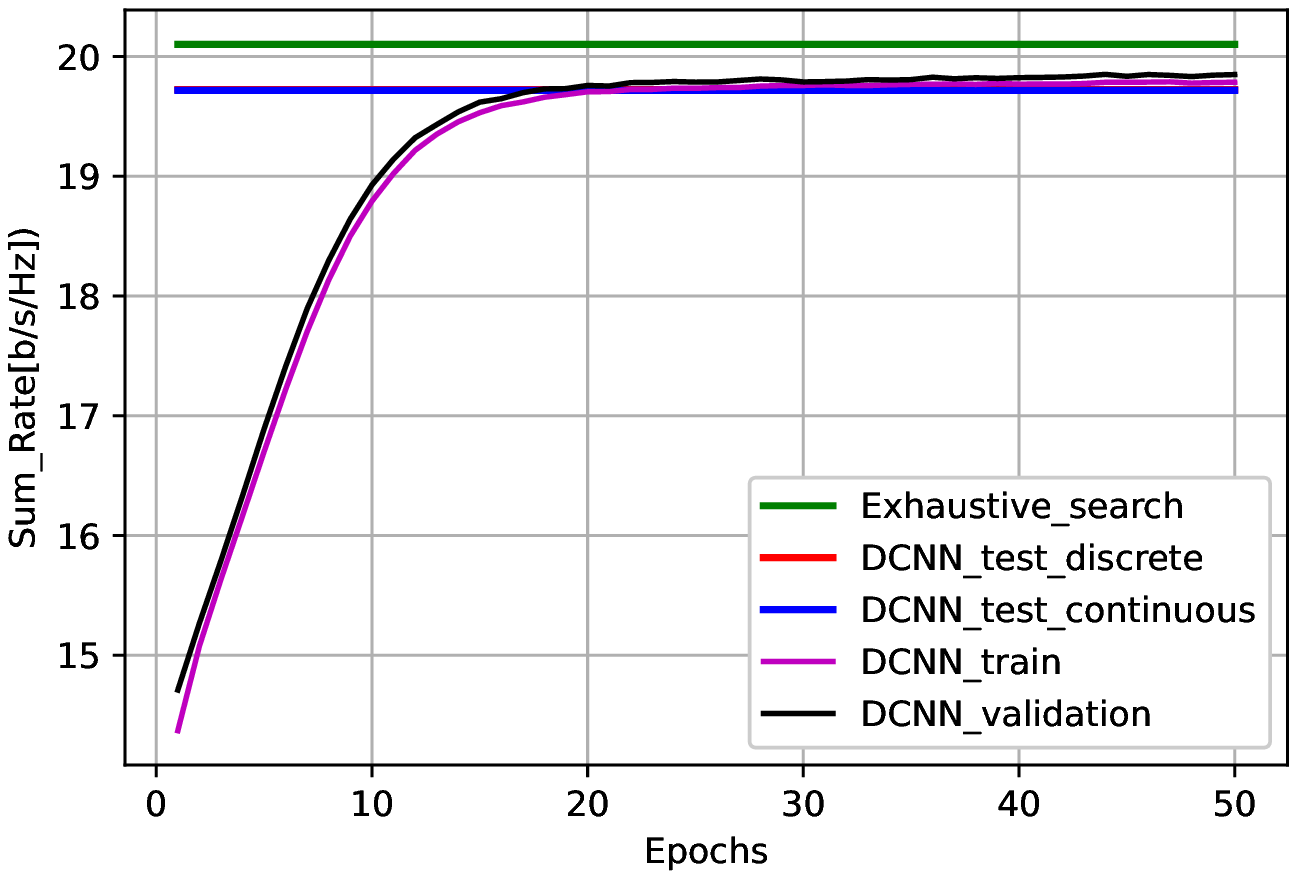}}
\caption{Training and testing DCNN unsupervised loss versus epochs and compared with exhaustive search method (a) LOS\_SU, (b) LOS\_MU}
\label{fig:singleUE}
\end{figure}

Fig. \ref{fig:singleUE} depicts the performance of both single-user and multi-user (two users) for LOS RA mode selection. The comparison shows DCNN training and validation performances regarding the network training part. The predicted loss via DCNN is denoted as DCNN\_test\_continuous, and the SE of the system is calculated via the predicted labels, which is plotted as DCNN\_test\_discrete. We observe that the predicted results almost reach the exhaustive search method.

\begin{figure}
\centering
\subfigure[]{\label{fig:a}\includegraphics[width=0.41\textwidth]{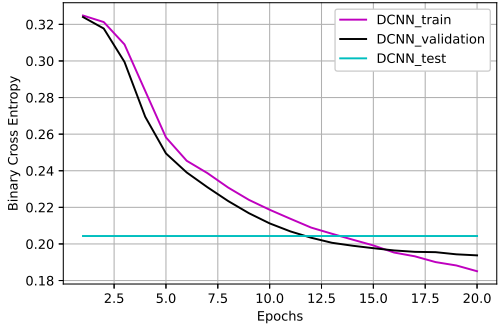}}
\subfigure[]{\label{fig:b}\includegraphics[width=0.41\textwidth]{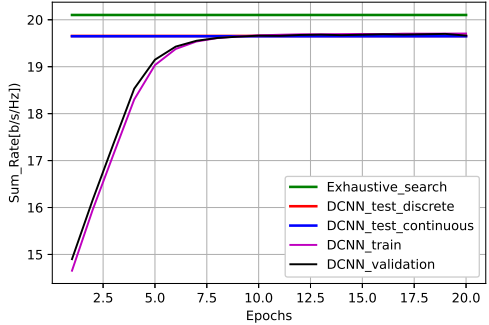}}
\caption{Training and testing DCNN semi-supervised loss versus epoch (a) supervised learning, (b) unsupervised learning for RA\_MU\_LOS}
\label{fig:unsup}
\end{figure}

As shown in Fig.\ref{fig:unsup}, the proposed DCNN semi-supervised learning for a two-user scenario can almost reach the optimum results, similar to the unsupervised one. However, here, we use half the sample size and fewer training epochs compared to the unsupervised DCNN. The proposed system model is similar to the unsupervised one, except that the number of epochs and batch sizes of the supervised step are $20$ and $128$, and for the unsupervised step are $20$ and $64$, respectively. Regarding the supervised DCNN, the stochastic gradient descent (SGD) optimizer with a $0.01$ learning rate and for the unsupervised DCNNs, the Adam optimizer with a learning rate of $0.0001$ is considered.

\subsection{Dynamic Scenario}
\begin{figure}
\centering
\subfigure[]{\label{fig:a}\includegraphics[width=0.24\textwidth]{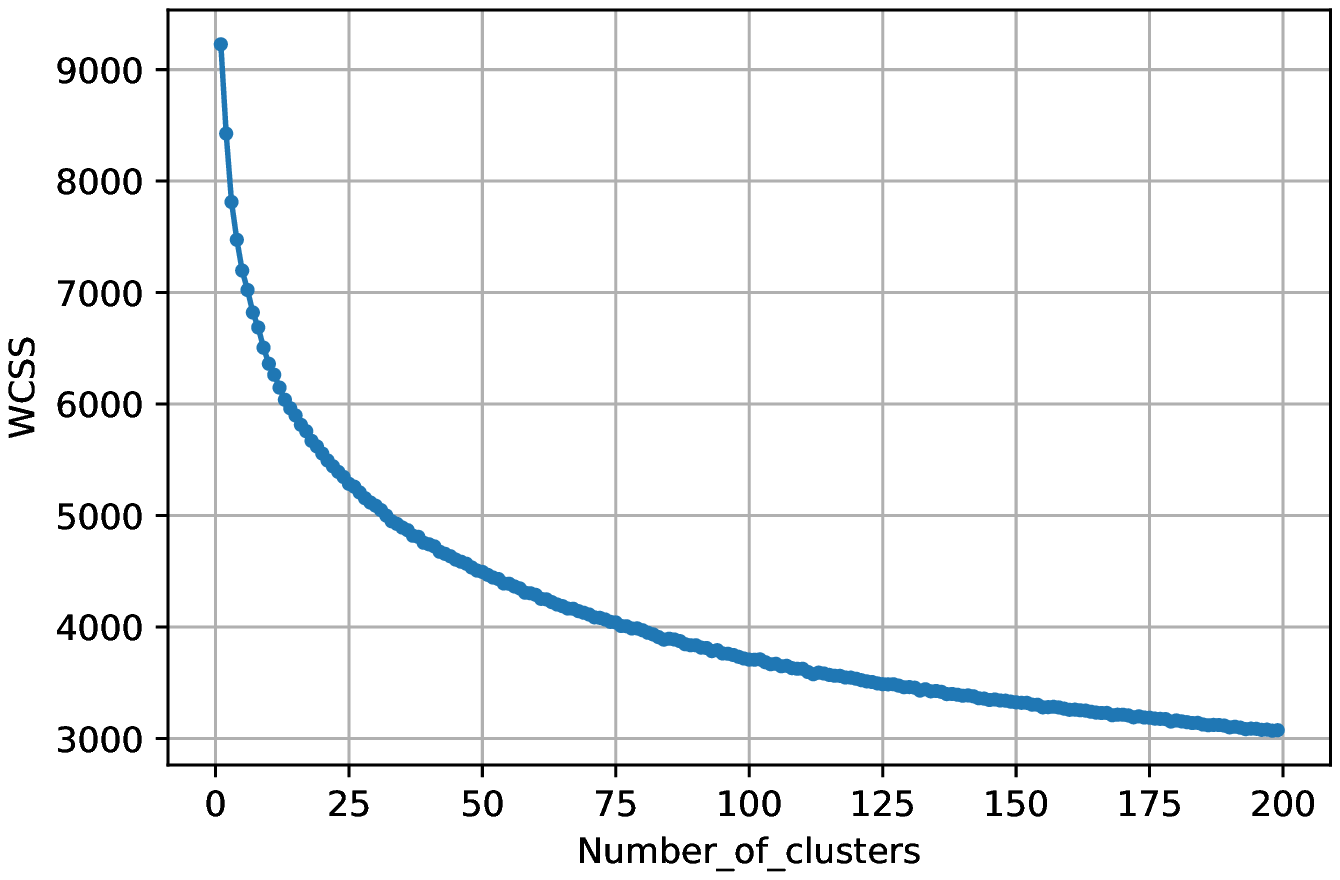}}
\subfigure[]{\label{fig:b}\includegraphics[width=0.24\textwidth]{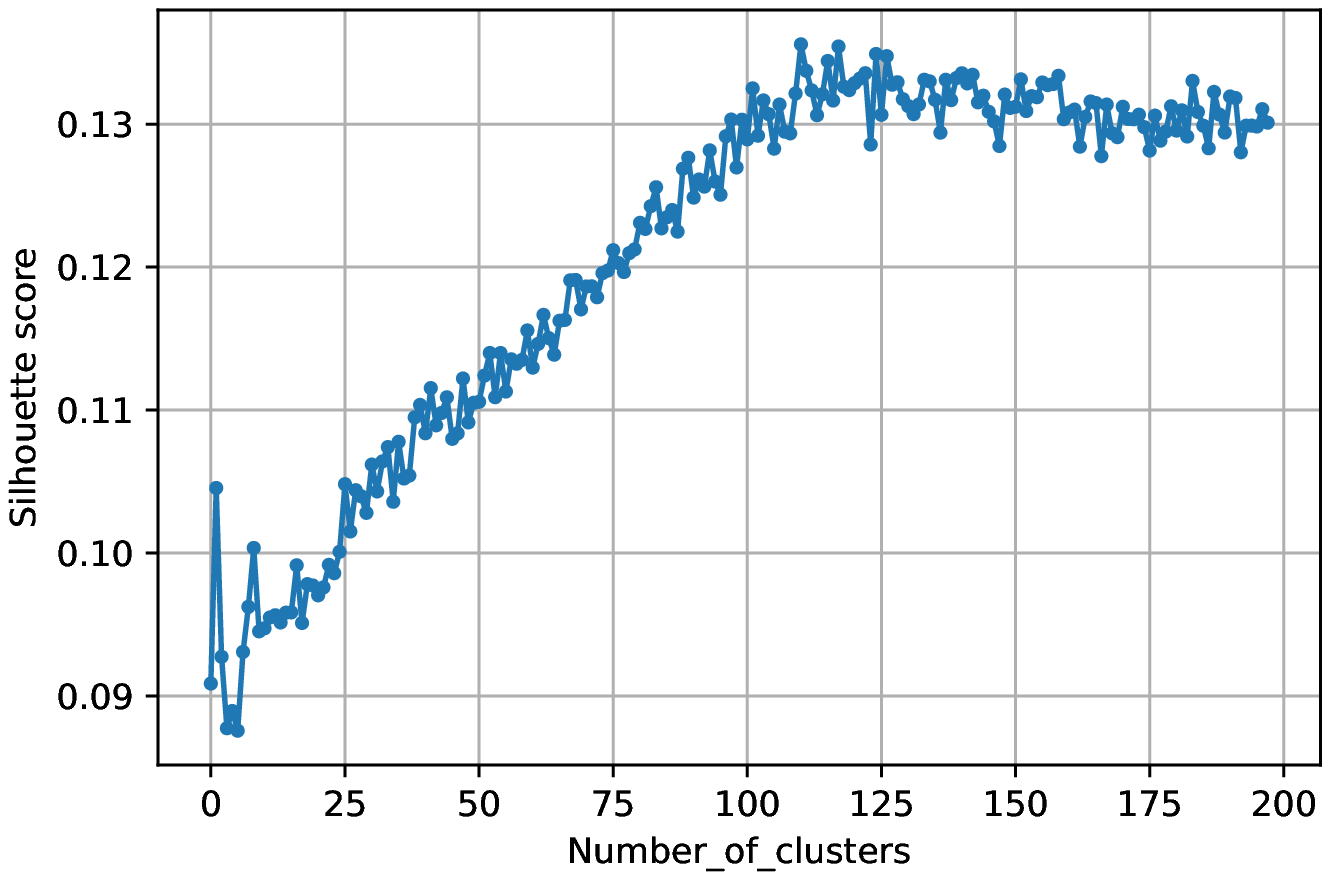}}
\caption{(a) Within sum of square distances (WCSS), (b) Silhouette metric for cluster number selection.}
\label{fig:clustering}
\end{figure}

\begin{figure}
\centering
\subfigure[]{\label{fig:MAB_Lineara}\includegraphics[width=0.41\textwidth]{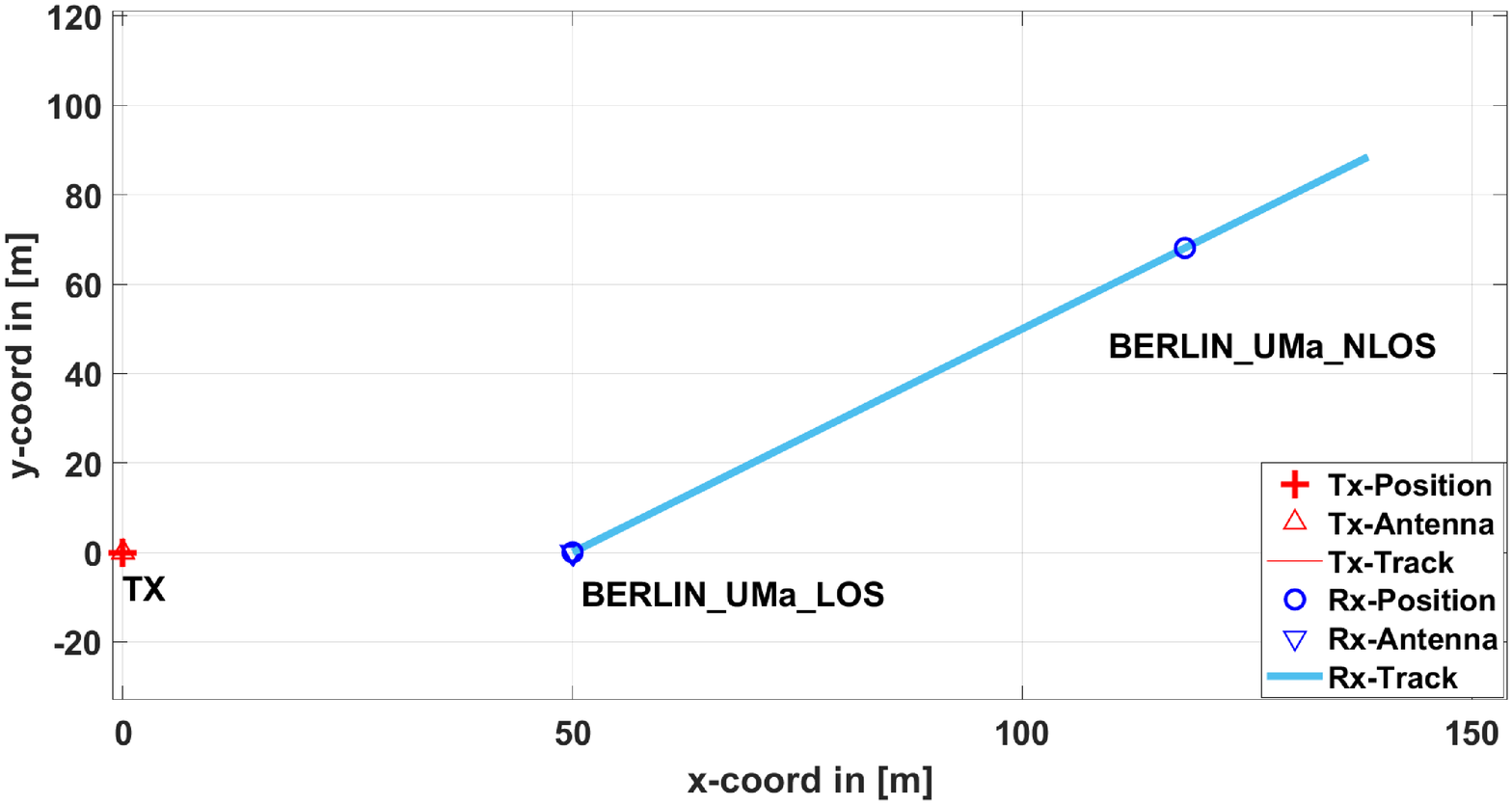}}
\subfigure[]{\label{fig:MAB_Linearb}\includegraphics[width=0.41\textwidth]{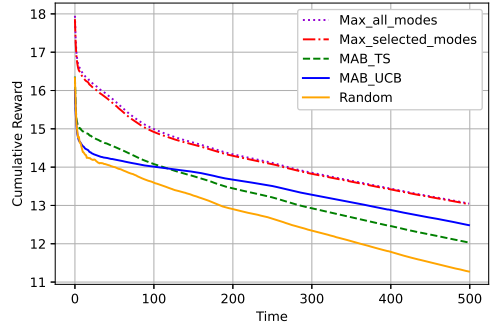}}
\caption{ (a) User trajectory, (b) Cumulative reward (SE) comparison for MAB methods, random selection, and exhaustive search.}
\label{fig:MAB_Linear}
\end{figure}

\begin{figure}
\centering
\subfigure[]{\label{fig:MAB_circa}\includegraphics[width=0.42\textwidth]{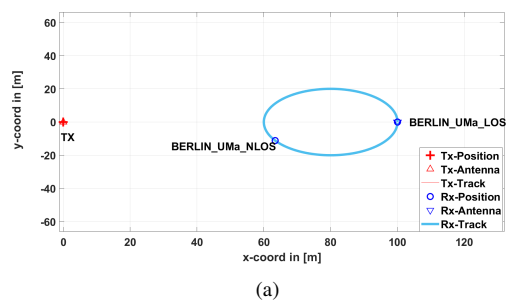}}
\subfigure[]{\label{fig:MAB_circb}\includegraphics[width=0.41\textwidth]{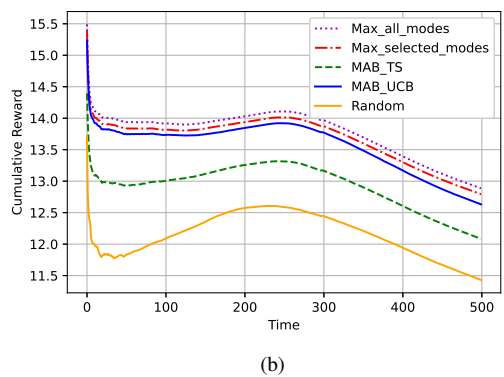}}
\caption{(a) User trajectory, (b) Cumulative reward (SE) comparison for MAB methods, random selection, and exhaustive search.}
\label{fig:MAB_circ}
\end{figure}

\begin{figure}
    \centering
    \includegraphics[width=0.41\textwidth]{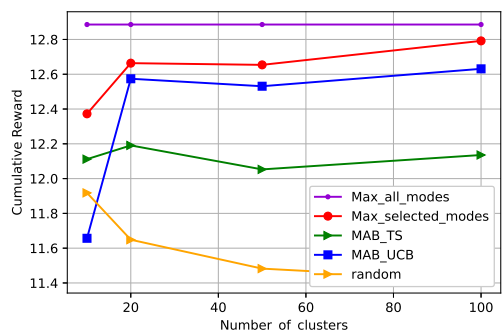}
    \caption{Cumulative reward (SE) versus the number of clusters}
    \label{fig:cluster}
\end{figure}

In the dynamic scenario, during offline training, we extract the PCA components ($100$ from $1600$ features) from the dataset. For finding a suitable number of clusters, we consider two metrics of WCSS and Silhouette \cite{thinsungnoena2015clustering}. As can be seen from Fig. \ref{fig:clustering}, $100$ clusters could separate RA states in a proper way.
Then, regarding the online training, two trajectories are considered for the UE. Fig. \ref{fig:MAB_Linear} depicts the result for a linear trajectory that comprises two segments with the UE moving at the speed of $30km/h$. For the MAB-UCB policy, we set $\gamma=5$. For the MAB-TS, the parameters are set as $\lambda=1$, $\omega=1$, $\Delta S = 10$, and $\Delta F = 50$. To assess dynamic RA state selection via MAB algorithms, we compare the result with random selection as a baseline in which, at each time step, RA mode is chosen from a uniform random distribution. As the upper bound, we consider both the exhaustive search method (max\_all\_modes) and the best state from the clustered representative data points (max\_selected\_modes), which maximizes the SE without considering the EE term. Fig. \ref{fig:MAB_Linearb} illustrates the cumulative reward (cumulative $\bar{R_i}$) comparison of this trajectory where the best of the selected states reaches the best of all states. In other words, $100$ clusters are sufficient to represent the states in this trajectory. In addition, MAB-UCB and MAB-TS are outperforming the random selection method, with the MAB-UCB performing better than the MAB-TS by increasing the time step.
In Fig. \ref{fig:MAB_circ}, a circular trajectory with a radius of $10m$ and a UE with a velocity of $30km/h$ is considered. As can be seen from the cumulative reward plot, the MAB-UCB in circular trajectory is close to the upper bound and outperforms the MAB-TS.
Moreover, Fig. \ref{fig:cluster} shows the effect of number of clusters on the MAB performance.

\section{Conclusion}
In this paper, we propose ML-based algorithms to solve the RA mode selection in both static and dynamic scenarios. In the static scenario, DCNNs for RA mode selection and digital beamforming design with both unsupervised and semi-supervised learning procedures are proposed in order to reduce the complexity of addressing the non-convex optimization problem. To facilitate these implementations, we customize the loss function and activation function of the network. In the case of RA mode selection in the dynamic scenario, employing offline training to extract the important features and clusters results in complexity reduction as well as improved performance. In addition, MAB policies are modified to fit the proposed approach.



\bibliographystyle{IEEEtran}
\bibliography{ref}

\end{document}